\documentclass[preprint,12pt]{elsarticle}
\biboptions{sort&compress}

\usepackage{amssymb}
\usepackage{amsmath}
\usepackage{hyperref}
\usepackage{float}
\usepackage{comment}

\begin{document}

\begin{frontmatter}



\title{PyOECP: A flexible open-source software library for estimating and modeling the complex permittivity based on the open-ended coaxial probe (OECP) technique}


\author{Tae Jun Yoon}
\author{Katie A. Maerzke}
\author{Robert P. Currier}
\author{Alp T. Findikoglu}
\address{Los Alamos National Laboratory, Los Alamos, NM 87545, the United States}

\begin{abstract}
We present PyOECP, a Python-based flexible open-source software for estimating and modeling the complex permittivity obtained from the open-ended coaxial probe (OECP) technique. The transformation of the measured reflection coefficient to complex permittivity is performed based on two different models, including the capacitance model and the antenna model. The software library contains the dielectric spectra of common reference liquids, which can be used to transform the reflection coefficient into the dielectric spectra. Several Python routines that are commonly employed (e.g., SciPy and NumPy) in the field of science and engineering are only required so that the users can alter the software structure depending on their needs. The modeling algorithm exploits the Markov Chain Monte Carlo method for the data regression. The discrete relaxation models can be built by a proper combination of well-known relaxation models. In addition to these models, the electrode polarization, which is a common measurement artifact for interpreting the dielectric spectra, can be incorporated in the modeling algorithm. A continuous relaxation model, which solves the Fredholm integral equation of the first kind (a mathematically ill-posed problem) is also included. This open-source software enables users to freely adjust the physical parameters so that they can obtain physical insight into their materials under test and will be consistently updated for more accurate measurement and interpretation of dielectric spectra in an automated manner. This work describes the theoretical and mathematical background of the software, lays out the workflow, validates the software functionality based on both synthetic and empirical data included in the software.\\

\end{abstract}


\end{frontmatter}


\section{Introduction}
\label{introduction}
	When a fluid is stimulated by an alternating current (AC) electric field, it shows a frequency-dependent response. From this electrical response, thermophysical properties, including specific conductance ($\kappa$), static dielectric constant ($\epsilon_\mathrm{r}$), and relaxation time ($\tau$), can be obtained. The excitation signal typically ranges from a few mHz to THz, depending on the measurement purpose and conditions. For instance, $\mathcal{O}(10^6)-\mathcal{O}(10^9)$ Hz signals are frequently employed for understanding the thermodynamic and kinetic aspects of small dipolar molecules, macromolecules, ions, and their intertwined molecular networks. These properties are important for both academic and industrial purposes. The relaxation behavior of dipoles in a polar liquid reveals the solvation structure (e.g., hydration number), structural relaxation process, and association behaviors of ions. The thermodynamic information from the relaxation process can thus be used for modeling the natural systems. The dielectric properties are also widely utilized for the in-line monitoring of physicochemical changes in industrially important systems. Some concrete examples are the in-line monitoring of chemical processes including separation and reaction \cite{kalamiotis2019optimised,bobowski2020permittivity,bur2002line,alie2004dielectric}, quality examination of agricultural products and foods in the field of agrophysics \cite{nelson2008dielectric}, medical diagnostics \cite{la2018open,gregory2020validation}, and monitoring of construction materials\cite{klein2004permittivity,chen2014broadband}.
	
	The electrical properties of a system are typically estimated based on two kinds of measurement methods: reflection and transmission. When a signal is exerted onto the material under test (MUT), one portion of the signal is reflected by the mismatch between the electrode and the MUT, whereas the other is transmitted. These transmitted or reflected signals are utilized in several measurement techniques, including the transmission line, cavity perturbation, multiple electrodes, and open-ended coaxial probe.
	
	Among them, the open-ended coaxial probe (OECP) technique is one of the most frequently used methods for the electromagnetic characterization of fluids based on the reflection method. In this method, a truncated (open-ended) coaxial line, which consists of an inner conductor, an outer conductor, and an insulator between the conductors, is inserted into a MUT. The excitation field from the generator connected to the probe propagates along the coaxial line and is reflected at the interface between the probe and the MUT. The characteristics of the reflected signal are often represented by a reflection coefficient $\Gamma(\omega)$, the ratio of the amplitude of the reflected signal to the incident one.
	\begin{equation}
		\Gamma(\omega)=\frac{Z_\mathrm{L}-Z_\mathrm{O}}{Z_\mathrm{L}+Z_\mathrm{O}}
	\end{equation}
	where $\omega$ is the angular frequency ($\omega=2\pi f$), and $Z_\mathrm{L}$ and $Z_\mathrm{O}$ are impedance (complex resistance) of a load and a reference, respectively. Compared to the other measurement techniques, the OECP technique is non-destructive, simple, and requires only a small amount of samples to acquire the reflection coefficient. For instance, it shows a better accuracy for estimating the dielectric constant compared to multiple electrode methods at high frequency.
	
	Taking into account these advantages, a variety of methods have been proposed to calibrate, measure, convert, and interpret the electrical properties based on the reflection coefficients. The measurement accuracy and stability could be improved by building measurement protocols (e.g., selection of proper calibration materials, probe geometry, conductor/insulator materials, sample containers, etc.) and stabilizing the measurement hardware (e.g., signal generator, filter, receiver, and amplifier). In spite of the aforementioned usefulness of the electrical property measurement and the advantages of the OECP technique, it is surprising that there is no open-source software or public library that contains numerical algorithms to convert the reflection coefficient into the complex permittivity spectra and to interpret them. The relative unpopularity of the measurement technique, compared to light spectroscopic techniques (typically higher than $\mathcal{O}(10^{12})$ Hz) and electrochemical impedance spectroscopy (typically less than $\mathcal{O}(10^7)$ Hz), would partially stem from the absence of such a library. Other reasons behind this unpopularity would be that (1) the measurement device (Vector Network Analyzer, VNA) can be costly and became commercially accessible only recently, and (2) the implementation of the numerical algorithms for the signal conversion and interpretation is not very easy as mentioned in previous works \cite{buchner2008can}.
	
	In this work, we would like to present our implementation of PyOECP, an open-source library that collects algorithms and necessary data to obtain and interpret the dielectric spectra obtained from the OECP technique. Currently, the software contains three algorithms for the conversion of reflection coefficient into complex permittivity, two types of regression algorithms to interpret the complex permittivity data and to correct the electrode polarization contribution automatically, and a property library that contains some dielectric spectra obtained from the literature and our measurements. The stability of the numerical algorithm and the data uncertainty were validated against two different VNAs and two different OECPs. In the following, the implementation and validation details are first offered. The results are briefly discussed, and other relevant features that have a high potential to be implemented in future will be discussed. 
	
	\section{Software overview}
	\subsection{Package, installation, and illustrations}
	PyOEPC is going to be available at the Los Alamos National Laboratory GitHub page (\url{GitHub.com/lanl}) and the developer’s GitHub page. It can also be installed via preferred installer program, which is included as one of the Python binary installers. The source code is written in Python 3. The software library exploits the \texttt{Numpy}, \texttt{Scipy} and \texttt{Matplotlib} library that are most commonly used in the field of science and engineering. The top directory of the package includes license file (\texttt{LICENSE}) and setup file (\texttt{setup}). In the directory \texttt{PyOECP}, three main modules (\texttt{Models}, \texttt{References}, and \texttt{Transform}) and \texttt{Examples} folder are included. In the Examples folder, four illustrative examples are included, which are also described in this work. Two of the demonstrations include experimentally measured reflection coefficient data, which are described in detail in Section 4. If a user is not familiar with the dielectric relaxation spectroscopy or open-ended coaxial probe technique, the scripts and data files in the folder \texttt{Examples} can be helpful.
	
	\subsection{Features}

	\begin{figure}[!ht]
		\begin{center}
		\includegraphics[width=90mm]{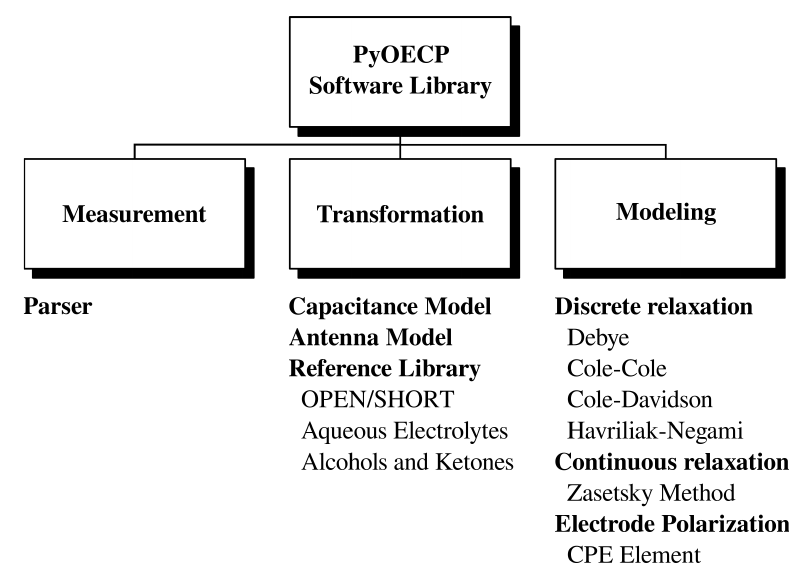}
		\caption{PyOECP software structure to obtain the dielectric spectrum from the reflection coefficient measurement. Three modules, including References, Transform, and Models are included for transforming the reflection coefficient into the dielectric spectra and for dissecting the spectrum based on relaxation models. The bold commands are the names of main functions (or classes) implemented in the software.}
		\label{fig1: software-structure}
		\end{center}
	\end{figure}

	Fig. \ref{fig1: software-structure} shows main functions in the software library, following a general procedure to estimate the dielectric spectrum of a MUT. After the measurement, users can obtain the complex reflection coefficient ($\Gamma=\Gamma'+i\Gamma''$) as a text file. This file is converted into a \texttt{NumPy} array. The resultant array contains the measurement frequency, real and imaginary components of the reflection coefficients as columns. It should be noted that this software does not offer pre-processing tools, such as the instrument calibration (e.g., port extension or OSL (Open/Short/Load) correction), and the experimental optimization of the measurement setup for enhancing the accuracy or stability. For instrument calibration and measurement setup, refer to the relevant reference articles \cite{la2018open,williams2003optimal,marks1996calibration}.

	After parsing the data, one of two numerical models ,capacitance model (function \texttt{capacitance}) and antenna model (function \texttt{antenna}), can be used to convert the reflection coefficients into the complex dielectric spectrum. The dielectric spectrum data can be decomposed and interpreted by fitting adequate models. For discrete relaxation models, the user can use the function \texttt{discrete} in conjunction with the implemented Markov Chain Monte Carlo method (class \texttt{MCMC}). To fit a continuous relaxation model, the user can use the function \texttt{continuous}. Selecting adequate models mainly depends on the nature of the MUT and the user’s interpretation, which will be dealt with to a certain extent in the illustrations. 
	
	There are several advantages in the PyOECP package. First, it combines the transformation and modeling parts, which are typically implemented independently. This enables users to work on their raw data in a consistent manner. The second strength of the package is its flexibility. That is, a user can add their custom functions or libraries easily to substitute the implemented functions. For instance, there have been a lot of efforts to utilize semi-empirical methods that can enable users to skip the calibration part and directly obtain the complex permittivity from the reflection coefficient. Since Python has a variety of packages for those purposes, they can be easily integrated with the package. In addition to the transformation part, the modeling part can also be modified easily. For instance, a user can replace the implemented MCMC algorithm with the complex nonlinear least square (CNLS) algorithm \cite{grosse2014program} depending on their preference.
	
	\subsection{Implementation details}
	\subsubsection{References}
	The reflection coefficient data measured by a VNA typically contains (1) frequency, (2) real part, and (3) imaginary part as a text file. The function \texttt{Parser} can be used to read the text file and transform it into a Numpy array. Currently, the function is written only for the text data generated from the Keysight Technology VNAs used in this work but can be easily customized to read any other formats by adjusting the number of header or tail lines.
	
	A proper choice of reference(s) in both models mainly depends on the characteristics of MUT or the reliability of the reference data. For instance, Kaatze examined the accuracy of the literature data of several pure substances, and showed that water is the most reliable substance that can be used as one of the references \cite{kaatze2007reference}. For weakly polar systems, short-chain alcohols (i.e., methanol, ethanol, 2-propanol, etc.) and ketones (i.e., acetone) are usually adopted as references. On the other hand, La Gioia et al. recently suggest that electrolyte solutions [e.g., NaCl (\textit{aq}) and KCl (\textit{aq})] as a reference material for biological samples \cite{la2018open}. In the currently released software, the data library \texttt{References} contains some common reference liquid data, which makes it easy for users to utilize them while converting the reflection coefficient. New data, which may be suitable for uncommon materials under test, can also be easily included by writing a custom function in the module. In the included demonstrations, we also validate that the reference liquids can influence the measurement accuracy.
	
	\subsubsection{Transform}
	There are a variety of methods to retrieve the frequency-dependent dielectric constant from the reflection coefficients. In classical models, the theory of the coaxial probe based on equivalent circuit representations is typically exploited. For instance, Marquardt suggested a transmission electromagnetic (TEM) mode model \cite{marquardt1963algorithm}, which shows a good performance for calculating the dielectric spectra. It is given as:
	\begin{subequations}
		\begin{equation}
		\begin{aligned}
			y=&\frac{k}{k_\mathrm{c}\ln(b/a)}\int_0^{\pi/2}\frac{\left[J_0(ka\sin\theta)-J_0(kb\sin\theta)\right]^2}{\sin\theta} d\theta\\&-\frac{i}{\pi}\frac{k}{k_\mathrm{c}\ln(b/a)}\int_0^{\pi}\left[2\mathrm{Si}\left(k\sqrt{a^2+b^2-2ab\cos\theta}\right)\right.\\&\left.-\mathrm{Si}\left[2ka\sin\left(\frac{\theta}{2}\right)\right]-\mathrm{Si}\left(2kb\sin\left(\frac{\theta}{2}\right)\right)\right]d\theta
		\end{aligned}
		\end{equation}
		\begin{equation}
			k=k_0\sqrt{\epsilon};\ k_\mathrm{c}=k_\mathrm{0}\sqrt{\epsilon_\mathrm{c}};\ k_0=\omega/c=2\pi f/c
		\end{equation}
		\begin{equation}
			\mathrm{Si}(x)=\int_0^x\frac{\sin t}{t}dt;\ J_0(x)=\sum_{n=1}^\infty\frac{(-1)^n}{2^{2n+1}(n!)^2}x^{2n}
		\end{equation}
		\label{eq:marquardt-equation}
	\end{subequations}

	In Eq. \ref{eq:marquardt-equation}, $y$ is the normalized input impedance ($y\equiv(1-\Gamma)/(1+\Gamma)$), $a$ and $b$ are the inner and outer radii of the conductor parts, $k_\mathrm{0}$ is the wavenumber in free space, $c$ is the speed of light, $\mathrm{Si}(x)$ is the sine integral, and $J_\mathrm{0}(x)$ is the Bessel function of the first kind. Since the direct calculation based on Eq. \ref{eq:marquardt-equation} is too extensive, several methods from the TEM mode model and its approximations to semi-empirical models to physics-guided neural network (PGNN) have been proposed to approximate the model while maintaining its accuracy \cite{stuchly1994new,athey1982measurement,marsland1987dielectric,gabriel1994admittance,komarov2016open,bartley1999permittivity,adedayo2014comparison,artioli2010particle}. Among them, we implemented capacitance model and antenna model, which are regarded as classical. 
	
	The capacitance model is one of the initial models used to estimate the dielectric constant \cite{stuchly1994new,stuchly1980coaxial,athey1982measurement}. In this model, three references of which the dielectric spectra at the measurement conditions are already known should be provided. Open (air, $\Gamma=1+0i$ ) and short ($\Gamma=-1+0i$) are frequently included as references ($i^2=-1$). The short can be measured by pressing the coaxial tip on a conducting material. The complex permittivity of the MUT from the capacitance model $\epsilon_\mathrm{MUT}^\mathrm{C}$ is then obtained as:
	\begin{equation}
		\epsilon_\mathrm{MUT}^\mathrm{C}=-\frac{\Gamma_\mathrm{MUT,2}\Gamma_\mathrm{1,3}}{\Gamma_\mathrm{MUT,1}\Gamma_\mathrm{3,2}}\epsilon_3
		-\frac{\Gamma_\mathrm{MUT,3}\Gamma_\mathrm{2,1}}{\Gamma_\mathrm{MUT,1}\Gamma_\mathrm{3,2}}\epsilon_2
		\label{eq:capacitance}
	\end{equation}
	Here, $\Gamma_\mathrm{i,j}$ and $\epsilon_\mathrm{k}$ ($i,j,k=1,\ 2,\ 3,$ and $\mathrm{MUT}$) is the reflection coefficient difference between the $i^\mathrm{th}$ and the $j^\mathrm{th}$ samples ($\Gamma_\mathrm{i,j}\equiv\Gamma_\mathrm{i}-\Gamma_\mathrm{j}$), and known complex permittivity of the $k^\mathrm{th}$ reference material, respectively. The complex permittivity of some common reference materials (e.g., NaCl (\textit{aq}), alcohols, and ketone) can be obtained from the module \texttt{References}. The implementation of the capacitance model is easy, and the calculation can be quickly done. However, it has been consistently noted that the obtained dielectric spectra is not satisfactorily accurate \cite{sasaki2018intercomparison,sasaki2015dielectric,ruvio2018comparison}. The main discrepancy arises from the absence of the radiation term, which makes it difficult to be used for the characterization of lossy materials. 
	
	The antenna model proposed by Marsland and Evans \cite{marsland1987dielectric} has shown a reasonable performance for a wide range of materials. This model includes the radiation effect in a lossy material as a first-order approximation term in the TEM mode model [Eq. \ref{eq:marquardt-equation}]. In the antenna model, four reference materials are chosen, and a short block measurement is typically included as the first reference material. The derived formula for obtaining the complex permittivity $\epsilon_\mathrm{MUT}^\mathrm{A}$ is given as:	
	\begin{equation}
		\begin{aligned}
		&G(\epsilon_\mathrm{MUT}^\mathrm{A})^{5/2}+\epsilon_\mathrm{MUT}^\mathrm{A}+\\&\frac{\Gamma_\mathrm{MUT,2}\Gamma_\mathrm{1,3}}{\Gamma_\mathrm{MUT,1}\Gamma_\mathrm{3,2}}(\epsilon_3+G\epsilon_3^{5/2})+\frac{\Gamma_\mathrm{MUT,3}\Gamma_\mathrm{2,1}}{\Gamma_\mathrm{MUT,1}\Gamma_\mathrm{3,2}}(\epsilon_2+G\epsilon_2^{5/2})=0
		\end{aligned}
		\label{eq:antenna}
	\end{equation}
	where the normalized radial conductance $G$ is obtained as:
	\begin{equation}
		G=\frac{-\Gamma_{4,1}\Gamma_{3,2}\epsilon_4+\Gamma_{4,2}\Gamma_{1,3}\epsilon_3+\Gamma_{4,3}\Gamma_{2,1}\epsilon_2}{\Gamma_{4,1}\Gamma_{3,2}\epsilon_4^{5/2}+\Gamma_{4,2}\Gamma_{1,3}\epsilon_3^{5/2}+\Gamma_{4,3}\Gamma_{2,1}\epsilon_2^{5/2}}
	\end{equation}
	Here, we let the first reference be the short measurement ($\Gamma_1=-1+0i$ and $\epsilon_1=\infty$). Since Eq. \ref{eq:antenna} is implicit for $\epsilon_\mathrm{MUT}^\mathrm{A}$, it should be numerically solved by iteration. As Marsland and Evans \cite{marsland1987dielectric} suggested, the complex permittivity obtained from the capacitance model is used as an initial estimate, and the Newton-Raphson algorithm is used for computing $\epsilon_\mathrm{MUT}^\mathrm{A}$.
	
	An additional functionality of the implemented module is to remove random noise in the measurement. The data regression step can be sensitive to the measurement noise, especially when the relaxation process occurs in the vicinity of the measurement frequency limit. After testing basic smoothing algorithms in the Scipy and NumPy library, we recommend to use the Savitzky–Golay filter \cite{savitzky1964smoothing} for signal processing. The default window length is set to be 81.
	
	Although the antenna model requires an additional reference, its calculation time is comparable to the capacitance model, and the accuracy is close to the most rigorous TEM mode model. We will compare the performance of two models in the following discussions by applying them to the experimental data. Considering both the calculation efficiency and accuracy, therefore, we recommend the antenna model as a default calculation method.
	
	\subsubsection{Models}
	In the module \texttt{Models}, the dielectric spectrum is modeled based on either discrete relaxation model(s) or a continuous relaxation time distribution for the identification of relaxation processes behind the spectra. The discrete relaxation model is represented as:\cite{kremer2002broadband}
	\begin{equation}
	\begin{aligned}
		\epsilon_\mathrm{MUT}&=\epsilon-i\epsilon''+\frac{\kappa}{i\omega\epsilon_0}\\&=\epsilon_\mathrm{\infty}+\sum_\mathrm{k=1}^\mathrm{N_\mathrm{m}}\frac{\Delta\epsilon_\mathrm{k}}{[1+(i\tau_\mathrm{k}\omega)^{1-\mathrm{\alpha_\mathrm{k}}}]^{1-\beta_\mathrm{k}}}+\frac{\kappa}{i\omega\epsilon_0}
	\end{aligned}
	\end{equation}
	where $\epsilon_\mathrm{MUT}$ is the complex permittivity of the MUT, $\epsilon_\mathrm{\infty}$ is the high-frequency dielectric constant, $i$ is the imaginary unit number, $\tau_\mathrm{k}$, $\Delta\epsilon_\mathrm{k}$, $\alpha_\mathrm{k}$ and $\beta_\mathrm{k}$ are the relaxation times, magnitudes, and exponents of the $k^\mathrm{th}$ relaxation mode among $N_\mathrm{m}$ numbers of relaxation processes, $\epsilon_0$ is the vacuum permittivity, and $\kappa$ is the specific conductance. The software user can build their own discrete relaxation models by either defining a custom relaxation function or super-positioning $N_\mathrm{m}$ relaxation models. The user can superposition more than two relaxation modes by using the \texttt{Parameter} class in the \texttt{Models} module. 
	
	When the MUT contains ionic species that can carry electric charge(s), the imaginary part of the dielectric spectra typically diverges as the measurement frequency decreases ($\omega\rightarrow0$). The divergence mainly originates from the formation of double layer at the tip of the coaxial probe, which is also known as the electrode polarization (EP).\cite{ishai2013electrode} Among many models for representing the EP contribution, we chose the model proposed by Bordi et al., which expresses the electrode polarization contribution as the constant phase element (CPE) \cite{bordi2001reduction}. According to this model, the influence of the electrode polarization can be modeled as:
	\begin{equation}
		\epsilon_\mathrm{meas}=\frac{\epsilon_\mathrm{MUT}}{1+A(i\omega)^{1-\gamma}\epsilon_\mathrm{MUT}}
	\end{equation}
	Thus, two additional parameters $A$ and $\gamma$ are required to remove the EP artifact. The CPE element is included as an input parameter in the function \texttt{Discrete}.
	
	The function \texttt{Discrete} is utilized to fit the dielectric spectra obtained from the module \texttt{Transform}. Two types of regression algorithms, CNLS and MCMC methods, are usually exploited. In the PyOECP, the basic MCMC method was implemented as a class \texttt{MCMC}. In the MCMC method, the initial parameters are first used to calculate the dielectric spectra $\epsilon^\mathrm{old}$. Next, one of the initial parameters is randomly chosen (function \texttt{Select}). The selected parameter $p$ is changed to $p'=p(1+D)$ where $D$ is a random number in a range of $[-R,R]$. Here, $R$ is a user-defined rate (function \texttt{Change}). The changed parameters are then used to compute a new dielectric spectra $\epsilon^\mathrm{new}$. The error (deviation) $\chi^2$  between the data to be fitted and the model spectra ($\epsilon^\mathrm{old}$ and $\epsilon^\mathrm{new}$) is defined as $\chi^2\equiv\sum_\mathrm{i}[(\epsilon_\mathrm{i}'-\epsilon_\mathrm{i,data}')^2+(\epsilon_\mathrm{i}''-\epsilon_\mathrm{i,data}'')^2]$ where $\epsilon_\mathrm{i}'$ and $\epsilon_\mathrm{i}''$ are the real and imaginary parts of the complex permittivity at the $i^\mathrm{th}$ frequency. If $\chi_\mathrm{new}^2$ is lower than $\chi_\mathrm{old}^2$, the change is accepted. That is, the parameter $p$ is changed to $p'$. Otherwise, a random number between 0 and 1 is generated and compared to the likelihood ratio, which is defined as:
	\begin{equation}
		L=\exp\left(\frac{-\chi_\mathrm{new}^2+\chi_\mathrm{old}^2}{2}\right)
	\end{equation}
	If the random number is between zero and $L$, the change is accepted even though it will increase the deviation. This acceptance step can be critical when there are a lot of local minima. If the random number is higher than $L$, the change is rejected. After either the acceptance or rejection, the algorithm checks if the number of the iteration reached the user-defined iteration number $N_\mathrm{iter}$. If $N<N_\mathrm{iter}$, a parameter is randomly selected again, and repeat the procedure. When the iteration number reaches the pre-defined maximum, the algorithm is terminated. Following this procedure, the algorithm tries to find an optimal parameter set that has the smallest deviation from the experimental data. 
	
	In a typical MCMC procedure, the chain is dissected into (1) a burn-in stage and (2) production stage. In the burn-in stage, the parameter set is changed without saving them so that the parameter set converges to an optimal solution whose error $\chi^2$ is close to its global maximum. In the production stage, all parameter sets changed at each step are saved for the error analysis. The default burn-in run length is set to be a quarter of the production run. The default maximum change rate (the variable \texttt{Rate} in the class \texttt{MCMC}) $R$ is set to be 0.01. The change rate parameter can affect the chain length for the convergence and the acceptance rate. In default, the function \texttt{Run} prints the acceptance rate, so that the user can control the maximum change rate. In the currently released software, the same change rate is applied for all parameters, but the change rate setting can be assigned for individual parameters. The change rate can also be adapted for matching the desired acceptance rate. 
	
	While the discrete relaxation models are widely utilized for understanding the underlying relaxation processes in a solution, a continuous relaxation time distribution model was also proposed and utilized for analyzing data. In this model, the dielectric spectrum is given as:
	\begin{equation}
		\frac{\epsilon-\epsilon_\infty}{\epsilon_\mathrm{t}-\epsilon_\infty}=\int_0^\infty\frac{g(\tau)}{1+i\omega\tau}d\tau
		\label{eq:zasetsky-buchner}
	\end{equation}
	where $\epsilon_\mathrm{t}$ is the dielectric constant at low frequency and $g(\tau)$ is the relaxation time distribution (probability density function). Eq. \ref{eq:zasetsky-buchner} is classified as the Fredholm equation of the first kind, which is a mathematically ill-posed problem. Several algorithms were proposed for obtaining the solution of the integral equation \cite{tuncer2001dielectric,malkin2006use,zasetsky2010quasi,schafer1996novel,macutkevic2004determination,bell1978solutions,yoon2020dielectric}. The PyOECP library uses the quasi-linear regression algorithm proposed by Zasetsky and Buchner \cite{zasetsky2010quasi}. This algorithm hypothesizes that a continuous relaxation distribution can be represented as a superposition of infinite (or many) numbers of Debye relaxation terms. For the algorithmic details, see the relevant articles \cite{tuncer2001dielectric,malkin2006use,zasetsky2010quasi,schafer1996novel,macutkevic2004determination,bell1978solutions,yoon2020dielectric}. 
	
	In the current implementation, the function Continuous requires the input to be $(\epsilon-\epsilon_\infty)/(\epsilon_\mathrm{t}-\epsilon_\infty)$. That is, the high-frequency dielectric constant $\epsilon_\infty$ should be determined before the use of the algorithm. $\epsilon_\infty$ can be determined by using the refractive index or by fitting an arbitrary discrete model to the measured data for removing the EP artifact and conductivity contribution.
		
	\section{Illustrations}
	In this section, we present several examples that validate and demonstrate the software capability. 
	
	\subsection{Data transformation}
	The first illustration uses both capacitance model and antenna model to convert the reflection coefficient into the complex permittivity. To validate the implementation of the module, we provide the reflection coefficient data obtained from two different VNAs and probes. One of the probes was an in-house made semi-rigid coaxial cable that was also successfully used in our previous work on the neodymium chloride solutions \cite{yoon2020dielectric}. This in-house made coaxial probe was connected to a VNA (Agilent PNA E8363C) of which the frequency domain was limited to 200 MHz$-$25 GHz. The other probe was purchased from Sequid GmBH (Model SDM-10G, Sequid GmBH). Since the working frequency range of the Sequid GmBH probe provided by the supplier was up to 10 GHz, this probe was connected to a low-frequency VNA unit (Agilent VBA E5061B) and used to measure the dielectric spectra in the frequency range from 10 MHz to 3 GHz. For the detailed information about the probes, refer to the relevant references \cite{yoon2020dielectric,wagner2011robust}.
	
	We selected methanol as the test material, considering that abundant data are available for the material. Methanol was purchased from Sigma Aldrich ($\geq99.9 \%$, Sigma Aldrich). Open, short, water, and acetone were used as references. For the short and water references, we used an aluminum foil and distilled water (Kroger) bought locally. Acetone was purchased from Sigma Aldrich ($\geq99.9 \%$, Sigma Aldrich). The measurement procedure was as follows. A solution (20 mL) was poured into a glass beaker and placed in a water bath (EW-208743-47, Cole-Parmer) for 30 minutes to maintain temperature at $25\pm0.1\ ^\circ$C. Then, the coaxial probes were dipped in the solutions to measure the reflection coefficients. All these reflection coefficient data are available in \texttt{Examples/Example1-Methanol/} folder.
	\begin{figure}[!ht]
		\begin{center}
		\includegraphics[width=90mm]{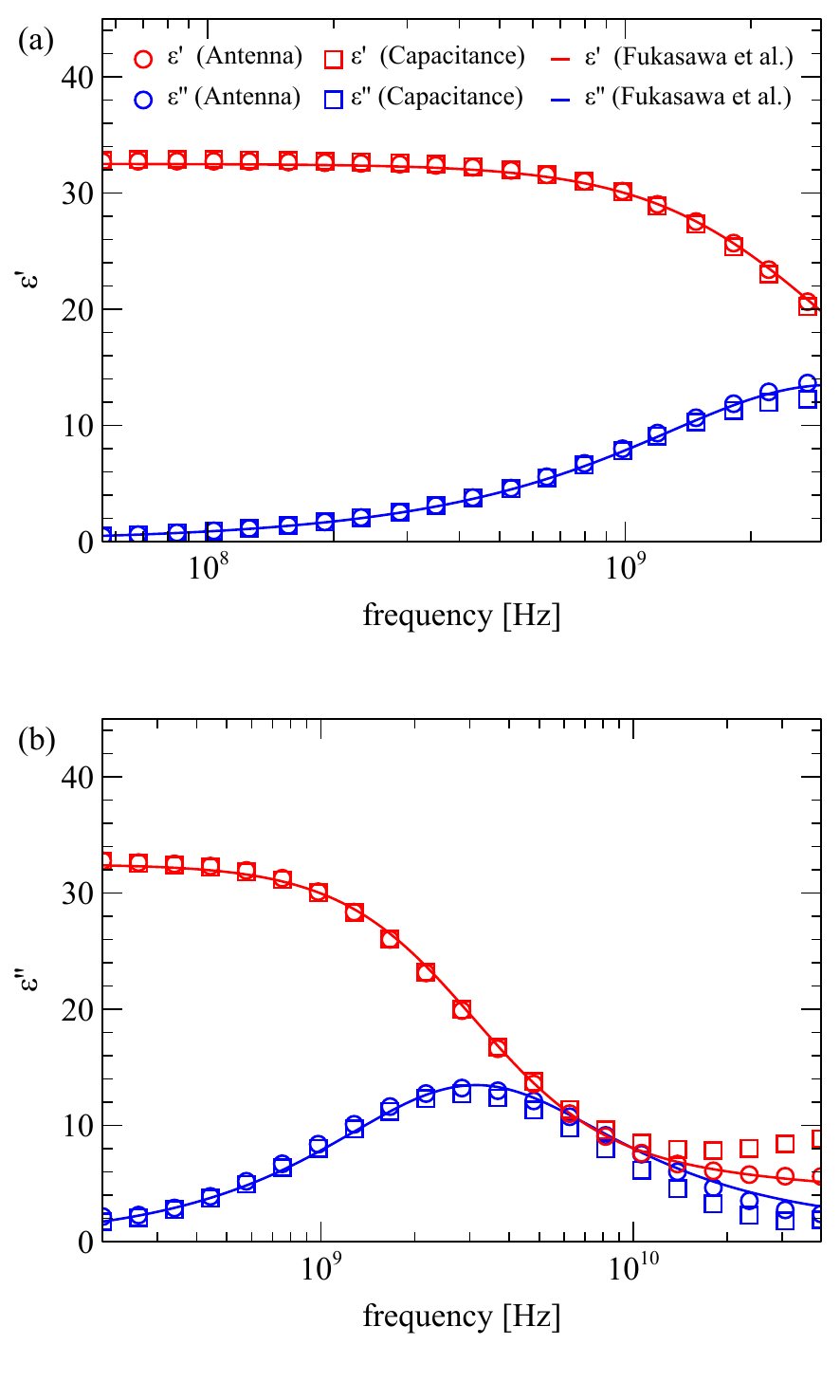}
		\caption{Dielectric spectra of methanol obtained from the module \texttt{Transform}. (a) Agilent PNA E8363C and in-house semirigid coaxial probe. (b) Agilent VBA E5061B and Sequid GmBH (Model SDM-10G, Sequid GmBH). Both functions \texttt{Capacitance} and \texttt{Antenna} yield reasonably good agreement with the literature data by Fukasawa et al. \cite{fukasawa2005relation} However, the high-frequency dielectric spectra from the capacitance model starts to deviate.}
		\label{fig2: methanol}
		\end{center}
	\end{figure}

	Fig. \ref{fig2: methanol} (a) and (b) shows the dielectric spectra of methanol obtained from different VNAs and sensors. The dielectric spectra of methanol from both functions \texttt{Capacitance} and \texttt{Antenna} in the module \texttt{Transform} show a reasonable consistency with the literature data. However, it is noteworthy that the deviation of the dielectric spectra from the capacitance model is larger than that from the antenna model, especially at high frequency. This result agrees well with the results obtained by Sasaki et al. \cite{sasaki2015dielectric} Therefore, we recommend users to utilize the antenna model, if it is not difficult to measure four references at the measurement conditions.
	
	\subsection{Modeling 1 (Discrete model without the electrode polarization)}
\begin{figure}[!ht]
		\begin{center}
		\includegraphics[width=90mm]{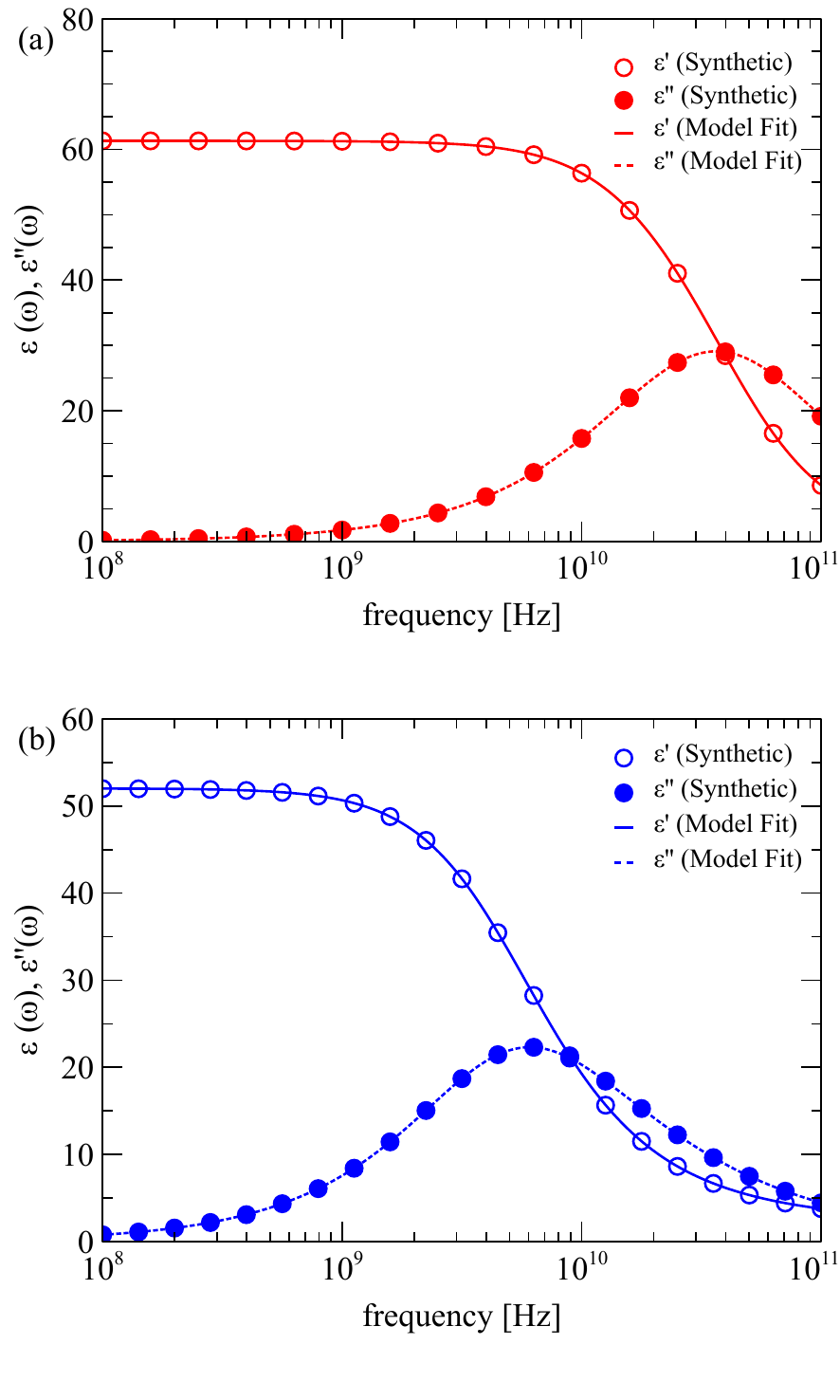}
		\caption{Comparison of the synthesized dielectric data and the fitting results. (a) two Debye relaxation model and (b) Havriliak-Negami model. Open and filled circles are real and imaginary parts of the synthetic data. The solid and dashed lines are the real and imaginary parts of the MCMC fitting results where the error function $\chi^2$ showed its minimum during the production run.}
		\label{fig3: synthetic-discrete}
		\end{center}
	\end{figure}

	The second case study attempts to perform a regression analysis for two synthetic datasets (two Debye relaxation model and Havriliak-Negami model). The first test dataset consists of two Debye relaxations. The parameters are given as $\epsilon_\infty=1.3$, $\Delta\epsilon=(50.0,\ 10.0)$  and $\tau=(3.979,\ 7.958)$. The other dataset (Havriliak-Negami model) was prepared by setting $\epsilon_\infty=2.0$, $\tau=3.183$, $\Delta\epsilon=50.0$, $\alpha=0.15$, and $\beta=0.08$. The function \texttt{Run} in the class \texttt{MCMC} (\texttt{Models} module) returns the parameter set of which the error $\chi^2$ was minimum, and a chain of parameter sets and $\chi^2$ obtained during the production run. Fig. \ref{fig3: synthetic-discrete}  (a) and (b) shows the fitting results for both synthetic datasets. The dielectric spectra of the synthetic data and the models agree well, and the parameters obtained show a reasonable agreement with the pre-determined parameters.
	
\begin{figure*}[!ht]
		\begin{center}
		\includegraphics[width=\textwidth]{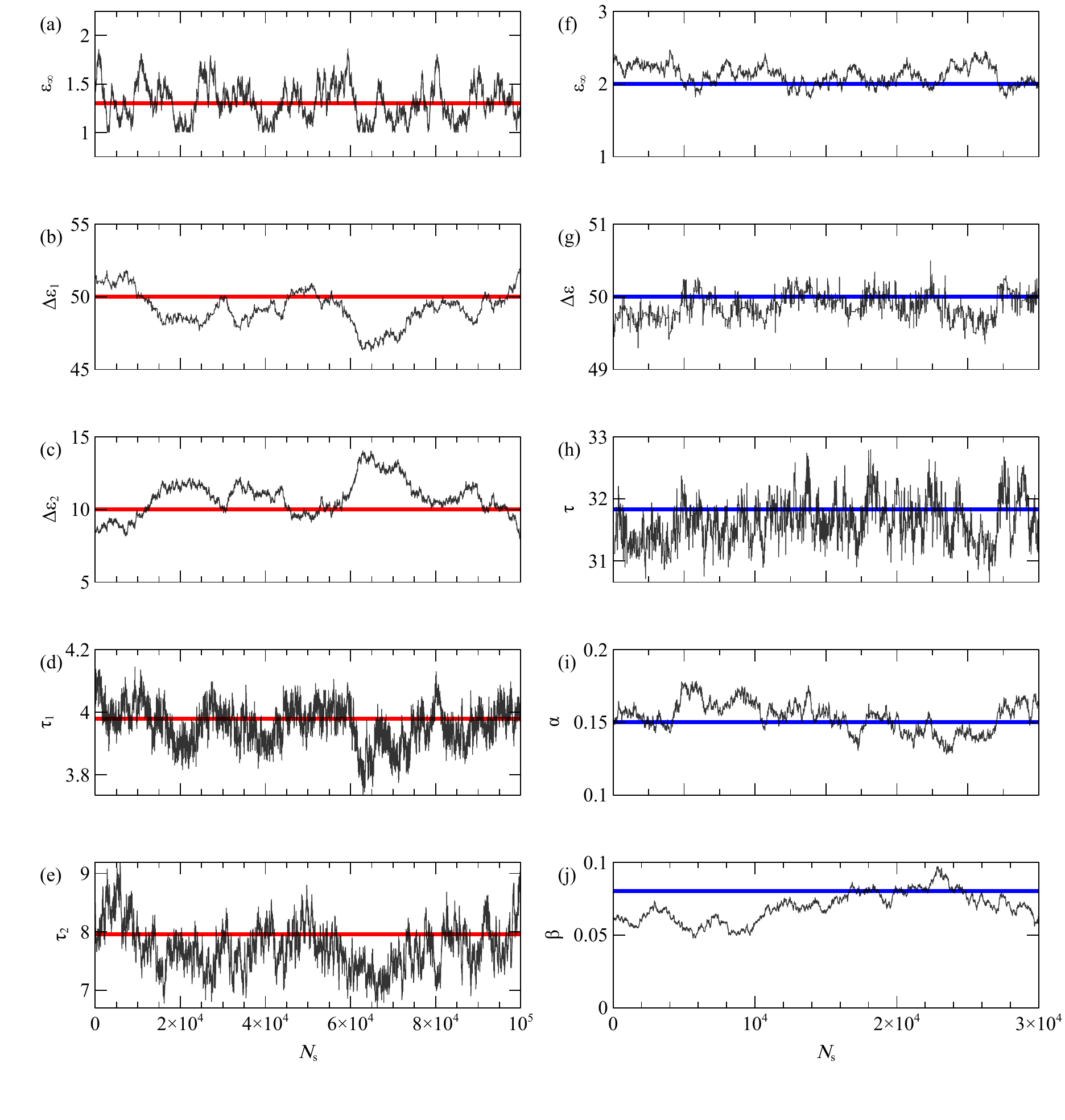}
		\caption{Changes in walkers (parameter set) visualized from the function Examine. (a) – (e) correspond to the parameter set of the two Debye relaxation model, and (f) – (j) correspond to that of the Havriliak-Negami model. The horizontal solid lines denote the exact parameters from the synthetic data.}
		\label{fig4: walkers}
		\end{center}
	\end{figure*}

	Some users might be interested in the statistical analysis. For instance, the $\chi^2$ information and the relevant model parameters in the production run can be used for estimating the data uncertainty. The function \texttt{Examine} in the class \texttt{MCMC} can be used to transform the chain data into a more convenient vector format and to visualize the fluctuations of each parameter (Fig. \ref{fig4: walkers}).
	
	\subsection{Modeling 2 (Discrete model with electrode polarization)}
	The goal of the third illustration is (1) to show the influence of references on the dielectric spectra and (2) to show the performance of the EP correction algorithm. To demonstrate the effect of reference liquids, the reflection coefficient data of NaCl (\textit{aq}) solutions at two different concentrations (0.09 mol/L and 0.18 mol/L) are obtained from the high-frequency VNA (Agilent PNA E8363C). NaCl solutions were prepared by adding NaCl salt ($\geq99.9 \%$, Sigma Aldrich) into a distilled water, and the measurement procedure (temperature control and contacting the sensor to the liquid media) was the same to the procedure described in Section 3.1.
	
	\begin{figure}[!ht]
		\begin{center}
		\includegraphics[width=90mm]{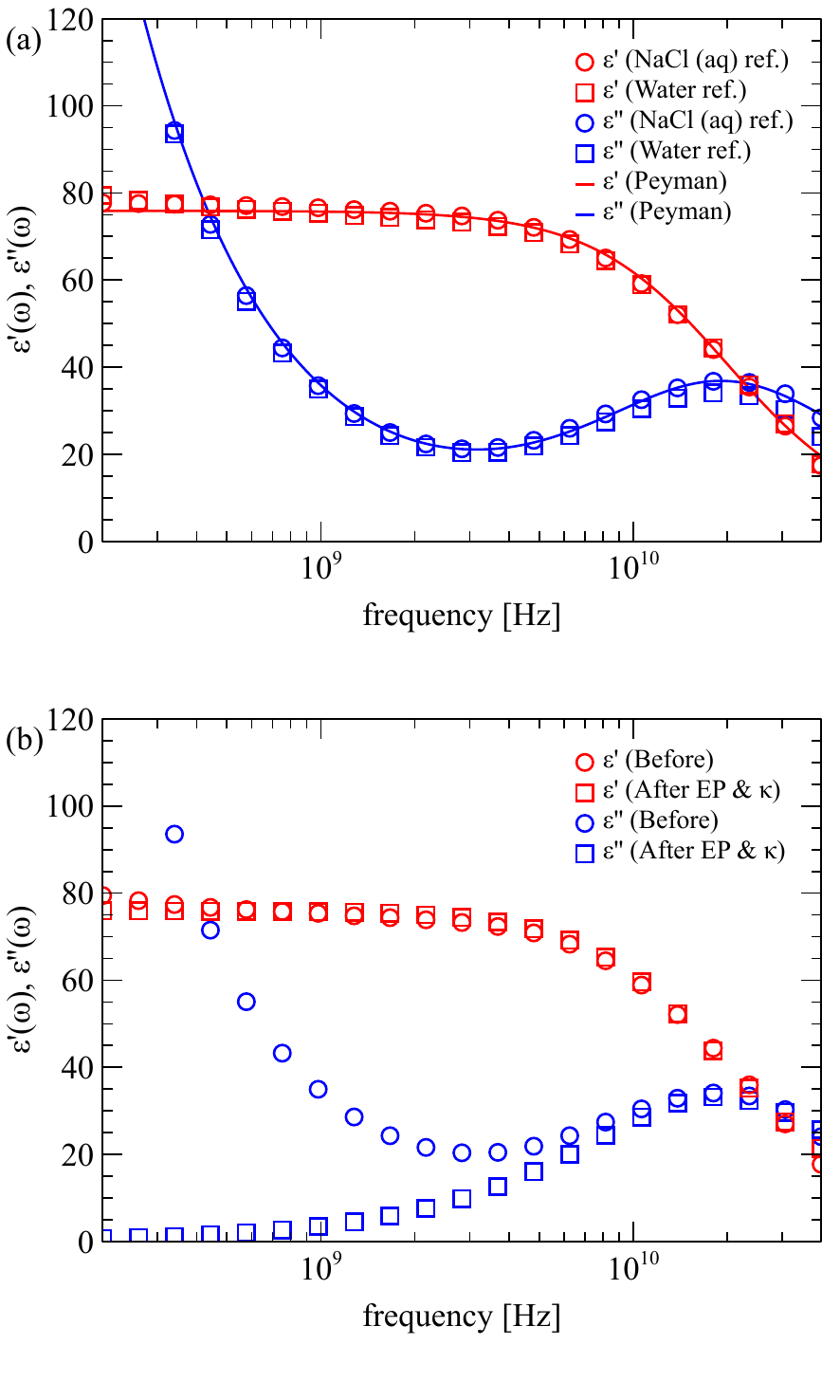}
		\caption{(a) Effect of references on the measurement accuracy of the dielectric spectra of NaCl (\textit{aq}) 0.18 mol/L. The open circles denote the dielectric spectrum obtained by using short, open, NaCl (\textit{aq}) 0.09 mol/L, and acetone as the references, whereas the open squares are obtained by using short, open, water, and acetone as the references. The dielectric spectrum obtained by using NaCl (\textit{aq}) 0.09 mol/L shows a better agreement with the literature data by Peyman et al. (solid lines). (b) Effect of the electrode polarization (EP) correction scheme by Bordi et al. for the dielectric spectrum obtained by using short, open, water, and acetone as the references. The dielectric spectrum became close to the literature data, but the high-frequency imaginary part still shows a slight deviation, which would come from the reference liquid effect.}
		\label{fig5: polarization}
		\end{center}
	\end{figure}
	Fig. \ref{fig5: polarization} (a) compares the dielectric spectra in NaCl (\textit{aq}) 0.18 mol/L solution obtained using different references. When short, open, water, and acetone were used as reference liquids for the function \texttt{Antenna}, the electrode polarization effect can be observed in the real part of the dielectric spectrum at the frequency range of 200 $-$ 300 MHz; the EP artifact makes the low-frequency real part of the dielectric spectrum diverge \cite{kremer2002broadband}. The imaginary part also diverges in the low-frequency domain, which arises from the specific conductivity contribution of ions. At high frequency, the imaginary part shows a slight deviation from the literature data provided by Peyman et al. \cite{peyman2007complex} On the other hand, when NaCl (\textit{aq}) 0.09 mol/L is used as reference instead of the distilled water, the high-frequency imaginary part of the dielectric spectrum shows a better agreement with the literature data. The EP artifact in the low-frequency real part is also slightly decreased since the literature data already takes into consideration the EP effect. Thus, a proper choice of the reference liquids can be crucial for increasing the measurement accuracy.
	
	Fig. \ref{fig5: polarization} (b) shows the effect of the EP correction algorithm implemented in this work. When the artifact is removed, the low-frequency dielectric spectrum of NaCl (\textit{aq}) 0.18 mol/L converges to the static dielectric constant. When both specific conductance contribution and EP contribution are removed in the imaginary part, the dielectric loss (imaginary part) converges to zero. Therefore, the implemented algorithm (the function CPE) can be utilized to remove the EP artifact.
	
	\subsection{Modeling 3 (Continuous relaxation model)}
	\begin{figure}
		\begin{center}
		\includegraphics[width=90mm]{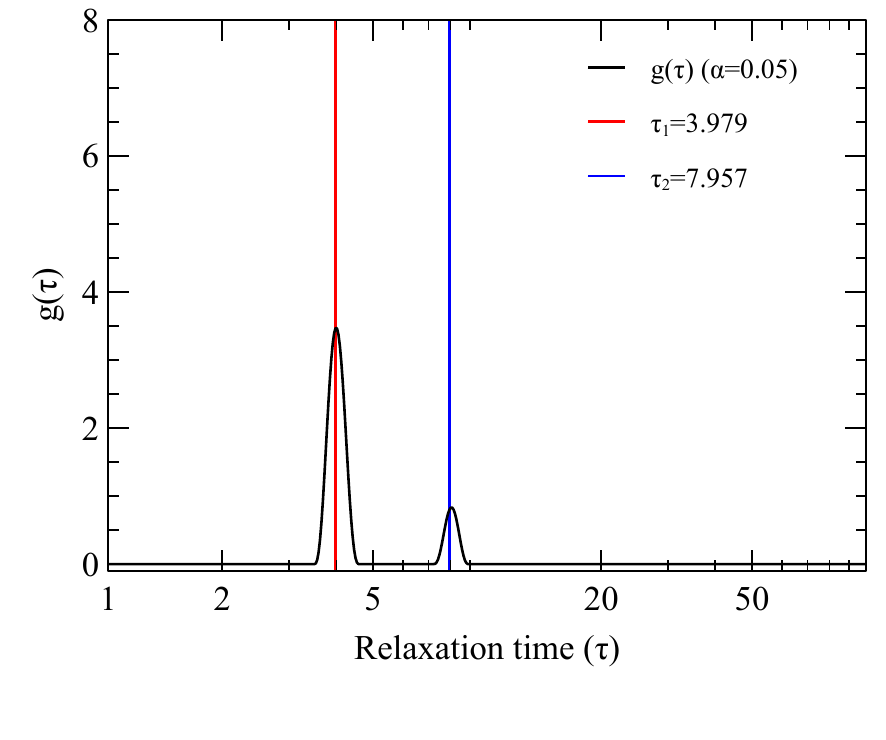}
		\caption{Comparison of the relaxation time distribution (black line) obtained from the function Continuous in the module Models and the relaxation time parameters (red and blue dashed lines) in the synthetic data (Two Debye model, $\epsilon_\infty=1.3$, $\Delta\epsilon_1=50.0$, $\Delta\epsilon=10.0$, $\tau_1=3.979$ and $\tau_2=7.957$). They agree well with each other.}
		\label{fig6: continuous}
		\end{center}
	\end{figure}
	The last example included in the package demonstrates the capability of the function \texttt{Continuous}. In this script, we use the two Debye relaxation model, which was synthesized in the \texttt{Example 2}. Fig. \ref{fig6: continuous} shows the continuous relaxation time distribution $g(\tau)$ obtained from the algorithm setting the smoothness parameter to be 0.05. The resultant modes in the distribution show a satisfactory agreement with the relaxation time in the synthetic dataset.
	
	\section{Conclusion}
	This work describes the guideline for installing and using the PyOECP, an open-source software library for measuring and modeling the dielectric spectra from the open-ended coaxial probe technique. PyOECP covers the whole standard dielectric spectrum measurement process (data-processing, data transformation, and interpretation parts). The software is organized in a flexible manner so that users can alter the subroutine depending on their preference, add additional functionalities, or utilize the submodules for other purposes. For instance, the modeling module can be directly used to fit the dielectric spectra from any sources (e.g., molecular dynamics (MD) simulations).
	
	We also demonstrate the performance and capability of the implemented functions (capacitance model, antenna model, Markov Chain Monte Carlo procedure for the regression based on discrete relaxation models, quasi-linear least square algorithm for the continuous relaxation model, electrode polarization correction algorithm) based on the synthetic data as well as the real-world reflection data we measured using different probes and vector network analyzers.
	
	We are planning to extend the capability of the PyOECP software library by including a variety of methods for the data transformation and modeling. For instance, the module \texttt{Transform} can be extended to include solvers for the TEM mode model, which are computationally expensive but known to be accurate, or the semi-empirical models such as the physics-guided neural network in conjunction with the finite element methods. For modeling part, the complex nonlinear least square algorithm or other models based on the fractional calculus can be implemented. Based on these research efforts, we hope that the disclosure and maintenance of the PyOECP can be helpful for scientists and engineers to utilize the OECP technique for its application to a variety of research topics.
	
	\section*{Declaration of competing interest}
	The authors declare that they have no competing financial interests or personal relationships that can influence this work.
	
	\section*{Acknowledgments}
	This work was supported in part by the Director’s Postdoctoral Fellow Program (20190653PRD4) as well as the Laboratory Directed Research and Development Program (20190057DR) at Los Alamos National Laboratory.




\bibliographystyle{elsarticle-num}
\bibliography{bibliography}
\end{document}